%% file: main.tex
\documentclass{llncs}

\usepackage{latexsym}

\input{macros}

\newcommand{\Update}[1]{
{\bf Update done here.}\\
}

\newcommand{\Polytype}{{\cal T}_{\Sort}}

\newcommand{\sgs}{{\cal T}_{\cal S}^{min}}

\newcommand{\gtchorpoX}[1]{\mathop{\gtpo}^{#1}}
\newcommand{\gechorpoX}[1]{\mathop{\gepo}^{#1}}
\newcommand{\gtchorpoXtype}[1]{\mathop{\gtpo}^{#1}_{\Type}}
\newcommand{\gechorpoXtype}[1]{\mathop{\gepo}^{#1}_{\Type}}

\newcommand{\newgeS}{\geS^{\ra}}
\newcommand{\newgtS}{\gtS^{\ra}}

\newcommand{\gtacc}{\gtsubt\!\!_{acc}}
\newcommand{\geacc}{\gesubt\!\!_{acc}}
\newcommand{\gepoacc}{\gtacc\!\!\succeq}

\newcommand{\gtchorpoXtypeacc}[1]{{\gepoacc}^{#1}_{\Type}}

\newcommand{\qgt}{\sqsupset}
\newcommand{\qge}{\sqsupseteq}

\newcommand{\judgF}{\turnstyle_{\!\!\!\!\!\Sigma}~}

\begin{document}

\title{\Large\bf HORPO with Computability Closure :\\ A Reconstruction}

\author{
  Fr\'ed\'eric Blanqui\inst{1}
  \and
  Jean-Pierre Jouannaud\inst{2}\thanks{Project
  LogiCal, P\^ole Commun de Recherche en Informatique du Plateau de
  Saclay, CNRS, \'Ecole Polytechnique, INRIA, Universit\'e Paris-Sud.}
  \and
  Albert Rubio\inst{3}}
 
\institute{
  {INRIA \& LORIA, Protheo team, Campus Scientifique, BP 239, 54506 Vand{\oe}uvre-l\`es-Nancy Cedex, France}
\and
  {LIX, \'{E}cole Polytechnique,
  91400 Palaiseau, France}
\and
  {Technical University of Catalonia,
  Pau Gargallo 5, 08028 Barcelona, Spain}  
}

\comment{
This work was partly supported by the RNTL
project AVERROES, France-Telecom, and the CICYT project LOGICTOOLS,
ref.\ TIN2004-07925.
}

\pagestyle{plain}
\bibliographystyle{plain}
\maketitle

\thispagestyle{empty}
\large

\begin{abstract}
  This paper provides a new, decidable definition of the higher-order
  recursive path ordering in which type comparisons are made only when
  needed, therefore eliminating the need for the computability
  closure, and bound variables are handled explicitly, making it
  possible to handle recursors for arbitrary strictly positive
  inductive types.
\end{abstract}

\input{introduction}

\input{preliminaries}
\input{ordering}

\input{normalization}
\input{conclusion}

\input{biblio}
\end{document}

%% file: macros.tex
\newcommand{\lambdax}{\lambda \hspace{-0.30mm} x}

\newcommand{\lambday}{\lambda \hspace{-0.55mm} y}

\newcommand{\Sort}{{\cal S}}

\newcommand{\Type}{{\cal T}_{\cal S}}
\newcommand{\gttype}{>_{\Type}}
\newcommand{\getype}{\geq_{\Type}}
\newcommand{\letype}{\leq_{\Type}}
\newcommand{\eqtype}{=_{\Type}}
\newcommand{\gtS}{\gttype}
\newcommand{\geS}{\getype}
\newcommand{\leS}{\letype}
\newcommand{\eqS}{\eqtype}

\newcommand{\Term}{\TT}

\newcommand{\gtF}{>_{\cal F}}

\newcommand{\geF}{\geq_{\cal F}}

\newcommand{\cand}[1]{[\![ #1 ]\!]}

\newcommand{\comment}[1]{}

\newcommand{\cqfd}{\hfill $\Box$}

\newcommand{\vect}[1]{\overline{#1}}

\newcommand{\turnstyle}{~\vdash~}

\newcommand{\ra}{\rightarrow}

\newcommand{\Dra}{\Rightarrow}

\newcommand{\Nat}{\mbox{$\mbox{I}\!\mbox{N}$}}

\newcommand{\TFX}{\TT(\F,\X)}

\newcommand{\TT}{{\cal T}}
\newcommand{\F}{{\cal F}}

\newcommand{\X}{{\cal X}}

\newcommand{\Var}[1]{{\cal V}ar({#1})}

\newcommand{\Dom}[1]{{\cal D}om({#1})}

\newcommand{\Ran}[1]{{\cal R}an({#1})}

\newcommand{\lrps}[2]{\mathop{\longrightarrow}^{#1}_{#2}}

\newcommand{\gtpo}{\succ}
\newcommand{\gepo}{\succeq}

\newcommand{\gtsubt}{\rhd}
\newcommand{\gesubt}{\unrhd}

\newcommand{\gehorpo}{\mathop{\gepo}_{horpo}}

\newenvironment{ruleset}
               {$$ \begin{array}{llcr}}{\end{array} $$ }

\newcommand{\Farc}[2]{\displaystyle \frac{#1}{#2}}

\newcommand{\newinfRULE}[3]
{\textbf{#1:}\\
\Farc{#2}{#3}}


%% file: introduction.tex
\section{Introduction}
\label{s:introduction}

The Higher-order Recursive Path ordering was first introduced
in~\cite{jouannaud99lics}. The goal was to provide a tool for showing
strong normalization of simply typed lambda calculi in which
higher-order constants were defined by higher-order recursive rules
using plain pattern matching.  Inspired by Dershowitz's recursive path
ordering for first-order terms, comparing two terms started by
comparing their types under a given congruence generated by equating
given basic types, before to proceed recursively on the structure of
the compared terms. In~\cite{jouannaud06jacm}, the type discipline was
generalized to a polymorphic type discipline with type constructors,
the congruence on types was replaced by a well-founded quasi-ordering
on types (in practice, a restriction of the recursive path ordering on
types), and the recursive definition itself could handle new
cases. There were two variants of the subterm case: in the first,
following the recursive path ordering tradition, a subterm of the
left-hand side was compared with the whole right-hand side; in the
second, a term belonging to the computability closure of the left-hand
side was used instead of a subterm. And indeed, a subterm is the basic
case of the computability closure construction, whose fixpoint
definition included various operations under which Tait and Girard's
notion of computability is closed. The ordering and the computational
closure definitions shared a lot in common, raising some expectations
for a simpler and yet more expressive definition able to handle
inductive types, as advocated in ~\cite{jouannaud06lpar}. This paper
meets these expectations (and goes indeed much further) with a new
definition of HORPO that improves over the previous
one~\cite{jouannaud06jacm} in several respects:

\begin{enumerate}
\item
There is a single decidable recursive definition, instead of a pair of
mutually inductive definitions for the computability closure and the
ordering itself;
\item
In contrast with the definition of HORPO with computability closure,
the new definition is decidable and syntax-directed (except, as usual,
for the subterm case);
\item
Type checking applies only when really needed, that is, when the
comparison does not follow from computability arguments;
\item
Bound variables are handled explicitly by the ordering, allowing for
arbitrary abstractions in the right-hand sides;
\item
Strictly positive inductive types are accommodated;
\item
There is no need for flattening applications on the right-hand side.
\end{enumerate}

This new definition appears to be powerful enough to prove strong
normalization of recursors for arbitrary strictly positive inductive
types.  The two major technical innovations which make it possible are
the integration of the computability closure within the ordering
definition on the one hand, and the explicit handling of binders on
the other hand. This integration of the computability closure is not
obtained by adding new cases in the definition, as was suggested
in~\cite{jouannaud06lpar}, but instead by eliminating from the
previous definition the unnecessary type checks.

%% file: preliminaries.tex
\section{Higher-Order Algebras}
\label{s:preliminaries}

Polymorphic higher-order algebras are introduced
in~\cite{jouannaud06jacm}. Their purpose is twofold: to define a
simple framework in which many-sorted algebra and typed
lambda-calculus coexist; to allow for polymorphic types for both
algebraic constants and lambda-calculus expressions. For the sake of
simplicity, we will restrict ourselves to monomorphic types in this
presentation, but allow us for polymorphic examples. Carrying out the
polymorphic case is no more difficult, but surely more painful.

Given a set $\Sort$ of {\em sort symbols} of a fixed arity, denoted by
$s:*^n\ra *$, the set of {\em types} is generated by the constructor
$\ra$ for {\em functional types}:

\[
\begin{array}{c}
\Polytype := s(\Polytype^n) ~|~ \Polytype \ra \Polytype\\
\mbox{for $s: *^n\ra *~\in\Sort$}
\end{array}
\] 

Types are {\em functional} when headed by the $\ra$ symbol, and {\em
data types} otherwise. $\ra$ associates to the right.  We use
$\sigma,\tau,\rho,\theta$ for arbitrary types.

Function symbols are meant to be algebraic operators equipped with a
fixed number $n$ of arguments (called the {\em arity}) of respective
types $\sigma_1,\ldots,\sigma_n$, and an
\emph{output type} $\sigma$. Let ${\F} = \biguplus_{\sigma_1,\ldots,\sigma_n,\sigma}
\F_{\sigma_1 \times\ldots \times \sigma_n \ra \sigma}$.
The membership of a given function symbol $f$ to $\F_{\sigma_1
\times\ldots \times \sigma_n \ra \sigma}$ is called a {\em type
declaration} and written $f : \sigma_1 \times\ldots \times \sigma_n
\ra \sigma$.

The set $\TFX$ of {\em raw algebraic $\lambda$-terms} is generated from
the signature ${\cal F}$ and a denumerable set $\X$ of
variables according to the grammar:
\[\Term := 
   \X ~|~ (\lambda \X : \Polytype . \Term) ~|~ @(\Term,\Term) ~|~ \F(\Term,\ldots,\Term).
\] 
The raw term $\lambdax:\sigma.u$ is an {\em abstraction} and $@(u,v)$
is an application.  We may omit $\sigma$ in $\lambdax:\sigma.u$ and
write $@(u,v_1,\ldots,v_n)$ or $u(v_1,\ldots,v_n)$, $n>0$, omitting
applications.  $\Var{t}$ is the set of free variables of $t$. A raw
term $t$ is \emph{ground} if $\Var{t}=\emptyset$.  The notation
$\vect{s}$ shall be ambiguously used for a list, a multiset, or a set
of raw terms $s_1,\ldots,s_n$.

Raw terms are identified with finite labeled trees by considering
$\lambdax :\sigma.u$, for each variable $x$ and type $\sigma$, as a
unary function symbol taking $u$ as argument to construct the raw term
$\lambdax : \sigma . u$.  {\em Positions} are strings of positive
integers.  $t|_p$ denotes the {\it subterm} of $t$ at position
$p$. We use $t\gesubt t|_p$ for the subterm relationship. The result
of replacing $t|_p$ at position $p$ in $t$ by $u$ is written $t[u]_p$.

An \emph{environment} $\Gamma$ is a finite set of pairs written as
$\{x_1:\sigma_1,\ldots,x_n:\sigma_n\}$, where $x_i$ is a variable,
$\sigma_i$ is a type, and $x_i\neq x_j$ for $i\neq j$. Our typing
judgements are written as $\Gamma \judgF s : \sigma$.  A raw term $s$
has type $\sigma$ in the environment $\Gamma$ if the judgement $\Gamma
\judgF s : \sigma$ is provable in the inference system given in
Figure~\ref{fig:terms}.  An important property of our type system is
that a raw term typable in a given environment has a unique type.

\begin{figure}
  \begin{center}
    \fbox{
      $
      \begin{array}{ccc}
	\begin{array}{c}
	  \newinfRULE
	      {Variables}      
	      {x : \sigma \in \Gamma}
	      {\Gamma \judgF x : \sigma}
	\end{array}
	&\quad &
	\begin{array}{c}
	  \newinfRULE
	      {Functions}
	      {\begin{array}{c}
		  ~f : \sigma_1 \times \ldots \times \sigma_n \ra \sigma \in \F\\
		  \Gamma \judgF t_1 : \sigma_1 ~\ldots~  \Gamma \judgF t_n :
		  \sigma_n
	      \end{array}}
	      {\Gamma \judgF f(t_1,\ldots,t_n): \sigma}
	\end{array}
	\\\\
	\begin{array}{c}
	  \newinfRULE
	      {Abstraction}
	      {\Gamma \cdot \{x : \sigma \} \judgF t : \tau}
	      {\Gamma \judgF (\lambda x:\sigma.t) : \sigma \ra \tau}
	\end{array}
	&\quad &
	\begin{array}{c}
	  \newinfRULE
	      {Application}
	      {\Gamma \judgF s : \sigma \ra \tau~~~~
		\Gamma \judgF t : \sigma }
	      {\Gamma \judgF @(s,t) : \tau}
	\end{array}
      \end{array}
      $
    }
  \end{center}
  \caption{The type system for monomorphic higher-order algebras}\label{fig:terms}
\end{figure}
Typable raw terms are called \emph{terms}. We categorize terms into
  three disjoint classes:
\begin{enumerate}
\item
\emph{Abstractions}, which are headed by $\lambda$;
\item
\emph{Prealgebraic}, which are headed by a function symbol, assuming
that the output type of $f\in\F$ is a base type;
\item
\emph{Neutral}, which are variables or headed by an application.
\end{enumerate}

\noindent
A \emph{substitution} $\sigma$ of domain
$\Dom{\sigma}=\{x_1,\ldots,x_n\}$ is a set of triples
$\sigma=\{\Gamma_1\judgF x_1\mapsto t_1,\ldots, \Gamma_n\judgF
x_n\mapsto t_n\}$, such that $x_i$ and $t_i$ have the same type in the
environment $\Gamma_i$. Substitutions are extended to terms by morphism,
variable capture being avoided by renaming bound variables when
necessary. We use post-fixed notation for substitution application.

A rewrite rule is a triple $\Gamma \judgF l\ra r$ such that
$\Var{r}\subseteq\Var{l}$, and $\Gamma\judgF l:\sigma$ and
$\Gamma\judgF r:\sigma$ for some type $\sigma$. Given a set of
rules $R$,
\[s\lrps{p}{l\ra r\in R} t \mbox{ iff } s|_p=l\gamma \mbox{ and }t=s[r\gamma]_p
\mbox{ for some substitution } \gamma\] 
The notation $l\ra r\in R$ assumes that the variables bound in $l,r$
(resp.  the variables free in $l,r$) are renamed away from the
free variables of $s[]_p$ (resp. the bound variables of $s[]_p$), to
avoid captures.

For simplicity, typing environments are omitted in the rest of the paper.

A {\em higher-order reduction ordering} $\succ$ is a well-founded ordering
of the set of typable terms which is

(i) \emph{monotonic}: $s \succ t$ implies that $u[s] \succ u[t]$;

(ii) \emph{stable}: $s \succ t$ implies that $s\gamma \succ t\gamma$ for all
 substitution $\gamma$.

(iii) \emph{functional}: $s\lrps{}{\beta}\cup\lrps{}{\eta}t$ implies $
 s \succ t$,

In~\cite{jouannaud06jacm}, we show that the rewrite relation generated
by $R\cup \{\beta,\eta\}$ can be proved by simply checking that $l>r$
for all $l\ra r\in R$ with some higher-order reduction ordering.

%% file: ordering.tex
\section{The Improved Higher-Order Recursive Path Ordering}
\label{ordering}

The improved higher-order recursive path ordering on higher-order
terms is generated from four basic ingredients: a \emph{type
ordering}; an \emph{accessibility} relationship; a \emph{precedence}
on functions symbols; and a \emph{status} for the function
symbols. Accessibility is a new ingredient originating from inductive
types, while the other three were already needed for defining HORPO.
We describe these ingredients before defining the improved
higher-order recursive path ordering.


\subsection{Ingredients}
\label{ssto}

\begin{itemize}
\item
A quasi-ordering on types $\geS$, called \emph{type ordering},
satisfying the following properties (let
${\gtS}={\geS\setminus\leS}$ be its strict part and
${\eqS}={\geS\cap\leS}$ be its associated equivalence relation):
\begin{enumerate}
\item
\label{wf}
\emph{Well-foundedness}: $\gtS$ is well-founded;

\item
\label{poa}
\emph{Arrow preservation}:
$\tau\ra\sigma \eqS \alpha~\mbox{iff}~
\alpha=\tau'\ra\sigma',~\tau'\eqS \tau~\mbox{and}~\sigma \eqS \sigma';$

\item
\label{sa}
\emph{Arrow decreasingness}:
$\tau\ra\sigma \gtS \alpha ~\mbox{implies}~ \sigma \geS \alpha \mbox{ or }
\alpha=\tau'\ra\sigma', \tau'\eqS \tau$ \mbox{ and $\sigma \gtS \sigma'$;}

\item
\label{ma}
\emph{Arrow monotonicity}:
$\tau\geS\sigma~\mbox{implies both}~ \alpha\ra\tau\geS\alpha\ra\sigma  ~\mbox{and}~
\tau\ra\alpha\geS\sigma\ra\alpha;$
\end{enumerate}

We denote by $\sgs$ the set of minimal types with respect to
${\newgeS}=({\gtS}\cup{\rhd})^*$ (reflexive and transitive closure).

We say that a  data type $\sigma$ occurs \emph{positively}
(resp. \emph{negatively}) in a type $\tau$ if $\tau$ is a data
type (resp. $\tau$ is a data type non equivalent to $\sigma$ in
$\eqS$), or if $\tau=\rho\ra\theta$ and $\sigma$ occurs positively
(resp. negatively) in $\theta$ and negatively (resp. positively) in
$\rho$.

\item
A set $Acc(f)$ of accessible arguments for every function declaration
$f:\sigma_1\ldots\sigma_n \ra \sigma$ with $\sigma$ being a data
type, where $i\in[1..n]$ is said to be \emph{accessible} if all data types
occuring in $\sigma_i$ are smaller than $\sigma$ in the
quasi-order $\getype$, and in case of equivalence (with $\eqtype$),
they must occur only positively in $\sigma_i$. Note that the application
operator $@:(\alpha\ra\beta)\times\alpha\ra\beta$ can be seen as a
function symbol with an empty set of accessible positions, since its
output type $\tau$ may occur negatively in any of its two argument
types $\sigma$ and $\sigma\ra\tau$.

A term $u$ is \emph{accessible} in $f(\vect{s})$, $f\in\F$,
iff there is $i\in Acc(f)$ such that $u=s_i$ or $u$ is \emph{accessible}
in $s_i$.
\emph{Accessibility} for $f\in\F\cup\{@\}$ is now
obtained by adding the minimal type subterms: $s\gtacc v:\tau$ iff $v$
is accessible in $s$, or $\tau\in\sgs$, $v$ is a strict subterm of $s$
and $\Var{v}\subseteq\Var{s}$. We denote by $\geacc$ the reflexive
closure of $\gtacc$.

\item
A \emph{precedence} $\geF$ on $\F\cup\{@\}$,
with $f\gtF @$ for all $f\in\F$.

\item
  A status (lexicographic or multiset) for all symbols in
  $\F\cup\{@\}$ with $@\in Mul$. The status of the symbol $f$ is
  denoted by $stat_f$.
\end{itemize}


We recall important properties of the type ordering~\cite{jouannaud06jacm}:

\begin{lemma}
\label{l:pdt}
Assuming $\sigma\eqS \tau$, $\sigma$ is a data type iff
$\tau$ is a data type.
\end{lemma}

\begin{lemma}
\label{l:typeord}
  Let $\geS$ be a quasi-ordering on types such that $\gtS$ is
  well-founded, arrow monotonic and arrow preserving.
 Then, the relation ${\newgeS}$, defined as ${({\geS}\cup{\rhd})^*}$, is a
  well-founded quasi-ordering on types extending $\geS$ and $\rhd$,
  whose equivalence coincides with $\eqS$.
\end{lemma}

\begin{lemma}
\label{l:non-emptyness}
$\sgs$ is a non-empty set of data types if $\Polytype\neq\emptyset$.
\end{lemma}


\subsection{Notations}

\begin{itemize}
\item
$s\gtchorpoX{X} t$ for the main ordering, with a finite set of
variables $X\subset\X$, with the convention that $X$ is omitted when empty;
\item
$s:\sigma\gtchorpoXtype{X} t:\tau$ for $s\gtchorpoX{X} t$ and $\sigma\getype\tau$;
\item
$s:\sigma\geacc\gechorpoXtype{X}t:\tau$ for $s\geacc w$ for some $w$ and
$@(w,\vect{x}):\sigma'\eqS\tau\gechorpoXtype{} t$ for some
$\vect{x}\in X$.
\end{itemize}


\subsection{Ordering Definition}

\begin{definition}
$s:\sigma\gtchorpoX{X}t:\tau \mbox{ iff}$ either:
\begin{enumerate}

\item
\label{funleft}
$s=f(\vect{s})$ with $f\in\F$ and either of
\begin{enumerate}
\item
\label{subt}
$s_i\geacc\gechorpoXtype{X}t$ for some $i$
\item
\label{stat}
$t=g(\vect{t}) \mbox{ with } f=_\F g\in\F$,
$s\gtchorpoX{X}\vect{t}$
and \mbox{$\vect{s} ({\gtchorpoXtype{}}\cup{\gtchorpoXtypeacc{X}})_{stat_f} \vect{t}$}
\item
\label{prec}
$t=g(\vect{t}) \mbox{ with } f\gtF g\in\F\cup\{@\}$ and $s\gtchorpoX{X}\vect{t}$
\end{enumerate}

\item
\label{appleft}
$s=@(u,v)$ and either of
\begin{enumerate}
\item
\label{appsubt}
$u\geacc\gechorpoXtype{X}t$ or $v\geacc\gechorpoXtype{X}t$
\item
\label{appstat}
$t=@(u',v')$ and $\{u,v\}(\gtchorpoXtype{X})_{mul} \{u',v'\}$
\item
\label{beta}
$u=\lambdax:\alpha.w$ and $w\{x\mapsto v\}\gechorpoX{X} t$
\end{enumerate}

\item
\label{absleft}
$s=\lambda x:\alpha.u$ and either of
\begin{enumerate}
\item
\label{abssubt}
$u\{x\mapsto z\}\gechorpoXtype{X} t$ for $z:\alpha$ fresh
\item
\label{absstat}
$t=\lambda y:\beta.v$, $\alpha\eqtype \!\beta$ and
$u\{x\mapsto z\}\!\gtchorpoX{X} \!v\{y\mapsto z\}$ \mbox{for $z:\beta$ fresh}
\item
\label{eta}
$u=@(v,x)$, $x\not\in\Var{v}$ and $v\gechorpoX{X}t$
\end{enumerate}

\item
\begin{enumerate}
\item
\label{arbitraryleft}
\label{var}
$s\not\in\X$ and $t\in X$
\item
\label{precabs}
\label{appabs}
$s\not\in\X, s\neq\lambdax:\alpha.u$,
$t=\lambday:\beta.w$ and $s\gtchorpoX{X\cup\{z\}}w\{y\mapsto z\}$ 
\mbox{for $z:\beta$ fresh}
\end{enumerate}

\end{enumerate}
\end{definition}

Our ordering definition comes in four parts, the first three dealing
with left-hand sides headed respectively by an algebraic symbol, the
application symbol and an abstraction, while the fourth factors out
those cases where the right-hand side is a previously bound variable or
an abstraction.

Cases~\ref{funleft} are very similar (up to type checks) to those of
Dershowitz's recursive path ordering with the subterm case~\ref{subt},
the status case~\ref{stat} and the precedence case~\ref{prec}. So are
Cases~\ref{appleft} and~\ref{absleft}. One difference is that there is
an additional case for handling respectively beta and eta. A more
substantial difference is that variable renaming has become explicit.

The major innovation of this new definition is the annotation of the
ordering by the set of variables $X$ that were originally bound in the
right-hand side term, but have become free by taking some subterm. This
allows rule~\ref{precabs} to pull out abstractions from the right-hand
side regardless of the left-hand side term, meaning that abstractions
are smallest in the precedence. Note that freed variables become
smaller than everything else but variables.

One may wonder why Case~\ref{stat} is so complicated: the reason is
that using recursively $\vect{s} ({\gtchorpoXtype{X}})_{stat_f}
\vect{t}$ would yield to lose strong normalization of the ordering.

We give now an example of use of this new definition with the
inductive type of Brouwer's ordinals, which constructor
$lim$ takes an infinite sequence of ordinals to build a new, limit
ordinal, hence admits a functional argument of type $\Nat\ra Ord$, in
which $Ord$ occurs positively.  As a consequence, the recursor admits
a much more complex structure than that of natural numbers, with an
explicit abstraction in the right-hand side of the rule for
$lim$.


\begin{example}
Brouwer's ordinals.

\noindent
$0 : Ord \quad\quad\quad S : Ord\Dra Ord \quad\quad\quad lim : (\Nat \ra Ord)\Dra Ord$\\
$n : \Nat \quad\quad\quad N : \Nat \quad\quad\quad F : \Nat \ra Ord$\\
$rec : Ord\times\alpha\times(Ord\!\ra\!\alpha\!\ra\!\alpha)
\times((\Nat\!\ra\!Ord)\!\ra\!(\Nat\!\ra\!\alpha)\!\ra\!\alpha)\Dra\alpha$
$$
\begin{array}{lll}
rec(0,U,V,W) & \ra & U \\
rec(s(N),U,V,W) & \ra & @(V, N, rec(N, U, V, W)) \\
rec(lim(F),U,V,W) & \ra & @(W, F, \lambda n . rec(@(F,n), U, V, W))
\end{array}
$$
Although the strong normalization of such rules is known to be
difficult to prove, it is checked automatically by our ordering.
We only show how the third rule is included in the ordering.
\begin{center}
$
rec(lim(F),U,V,W) \gtchorpoXtype{} @(W, F, \lambda n . rec(@(F,n), U,
V, W))
$ 
\end{center}
yields 2 subgoals according to Case~\ref{prec}: $\alpha \getype \alpha$ and \\
$rec(lim(F),U,V,W) \gtchorpoX{} \{W, F, \lambda n.rec(@(F,n), U, V, W)\}$. \\
The first one is trivial and the second one simplifies to:
\begin{enumerate}
\item
$rec(lim(F),U,V,W) \gtchorpoX{} W$ which succeeds by Case~\ref{subt},
\item
$rec(lim(F),U,V,W) \gtchorpoX{} F$, which succeeds by
Case~\ref{subt} since $F$ is accessible in $lim(F)$,
\item
$rec(lim(F),U,V,W) \gtchorpoX{} \lambda n.rec(@(F,n), U, V, W)$ which yields, by Case~\ref{precabs}, 
$rec(lim(F),U,V,W) \gtchorpoX{\{n\}}\! rec(@(F,n), U, V, W)$ yielding, by Case~\ref{stat}, two goals
\begin{enumerate}
\item
$\{lim(F), U, V, W\} (\gtchorpoXtype{}\cup\gtchorpoXtypeacc{\{n\}})_{mul} \{@(F,n),U,V,W\}$, \\
which reduces to
$lim(F) \gtchorpoXtypeacc{\{n\}} @(F,n)$ which holds by
Case~\ref{subt} since $F$ is accessible in $lim(F)$, and
\item
$rec(lim(F),U,V,W)\! \gtchorpoX{\{n\}}\{@(F,n),U,V,W\}$, that decomposes 
into three goals trivially solved by Case~\ref{subt}, and \\
$rec(lim(F),U,V,W) \gtchorpoX{\{n\}} @(F,n)$ yielding, by Case~\ref{prec}, 
\begin{enumerate}
\item
$rec(lim(F),U,V,W) \gtchorpoX{\{n\}} F$, which holds by Case\ref{subt}, since $F$ is accessible in $lim(F)$, and
\item
$rec(lim(F),U,V,W) \gtchorpoX{\{n\}} n$ which holds by
Case~\ref{var}, therefore ending the computation.
\end{enumerate}
\end{enumerate}
\end{enumerate}
\end{example}

%% file: normalization.tex
\section{Strong Normalization}
\label{s:normalization}

\begin{theorem}
\label{t:phorpo}
$\gtchorpoXtype{+}$ is a decidable higher-order
reduction ordering. 
\end{theorem}

Contrasting with our previous proposal made of an ordering part and a
computability closure part, our new ordering is a decidable inductive
definition: $s\gtchorpoX{X}t$ is defined by induction on the triple
$(n,s,t)$, using the order
$(>_{\Nat},\lrps{}{\beta}\cup\gtsubt,\gtsubt)_{lex}$, where $n$ is
the number of abstractions in $t$. The quadratic time decidability
follows since all operations used are clearly decidable in linear
time.  The fact that $\gtchorpoX{X}$ is quadratic comes from those
cases that recursively compare one side with each subterm of the other
side.  This assumes of course that precedence and statuses are given,
since inferring them yields NP-completeness as is well-known for the
recursive path ordering on first-order terms. 

The stability and monotonicity proofs are routine. As the old one, the
new definition is not transitive, but this is now essentially due to
the beta-reduction case~\ref{beta}.  We are left with strong
normalization, and proceed as in~\cite{jouannaud06jacm}. The
computability predicate differs however in case of data types, since
it has to care about inductive type definitions.


\subsection{Candidate Terms}

Because our strong normalization proof is based on Tait and Girard's
reducibility technique, we need to associate to each type $\sigma$,
actually to the equivalence class of $\sigma$ modulo $\eqS$, a set of
terms $\cand{\sigma}$ closed under term formation.  In particular, if
$s\in\cand{\sigma\ra\tau}$ and $t\in\cand{\sigma}$, then the raw term
$@(s,t)$ must belong to the set $\cand{\tau}$ even if it is not
typable, which may arise in case $t$ does not have type $\sigma$ but
$\sigma'\eqS\sigma$. Relaxing the type system to type terms up to type
equivalence $\eqS$ is routine~\cite{jouannaud06jacm}. We use the
notation $t:_C\sigma$ to indicate that the raw term $t$, called a {\em
candidate term} (or simply, a term), has type $\sigma$ in the relaxed
system.

\subsection{Candidate Interpretations}

In the coming sections, we consider the well-foundedness of the strict
ordering $(\gtchorpoXtype{})^+$, that is, equivalently, the strong
normalization of the rewrite relation defined by the rules
$s\lrps{}{}t$ such that $s\gtchorpoXtype{} t$. Note that the set $X$
of previously bound variables is empty. We indeed have failed proving
that the ordering $(\gtchorpoXtype{X})^+$ is well-founded for an
arbitrary $X$, and we think that it \emph{is not}. As usual in this
context, we use Tait and Girard's computability predicate method, with
a definition of computability for candidate terms inspired
from~\cite{jouannaud06jacm,blanqui06rr}.

\begin{definition} 
\label{d:computability}
The family of {\em candidate interpretations}
$\{\cand{\sigma}\}_{\sigma\in\Type}$ is a family of subsets of the set of
candidate terms which elements are the least sets satisfying the following
properties:

(i) If $\sigma$ is a data type and $s:_C\sigma$ is neutral,
then $s\in\cand{\sigma}$ iff $t\in\cand{\tau}$ for all terms $t$
such that $s\gtchorpoXtype{} t:_C\tau$;

(ii) If $\sigma$ is a data type and $s=f(\vect{s}):_C\sigma$ is
prealgebraic with $f:\sigma_1\ldots\sigma_n\Dra\sigma'\in\F$, then
$s\in\cand{\sigma}$ iff $s_i\in\cand{\sigma_i}$ for all $i\in Acc(f)$
and $t\in\cand{\tau}$ for all terms $t$ such that $s\gtchorpoXtype{}
t:_C\tau$;

(iii) If $\sigma$ is the functional type $\rho\ra\tau$ then
$s\in\cand{\sigma}$ iff $@(s,t)\in\cand{\tau}$ for all
$t\in\cand{\rho}$;

A candidate term $s$ of type $\sigma$ is said to be {\em computable}
if $s\in\cand{\sigma}$. A vector $\vect{s}$ of terms of type
$\vect{\sigma}$ is computable iff so are all its components. A
(candidate) term substitution $\gamma$ is computable if all candidate
terms in $\{x\gamma \mid x\in\Dom{\gamma}\}$ are computable.
\end{definition} 

Our definition of candidate interpretations is based on a
lexicographic combination of an induction on the well-founded type
ordering $\newgtS$, and a fixpoint computation for data
types. \comment{This is so since

(i) the type of the right-hand side term strictly 
decreases in Case~\ref{precabs}: let $s:\sigma$ and
$u\{y:\beta\mapsto z:\beta\}:\tau$ be the terms compared in
Case~\ref{precabs}, and assume that
$s:\sigma\gtchorpoXtype{X}t=\lambday:\beta:u$ is the originating
comparison, hence $\sigma\geS\beta\ra\tau$; by Lemma~\ref{l:typeord},
we get $\sigma\gtS\tau$, showing our claim;

(ii) the type of the right-hand side term has not increased in
Case~\ref{var}, thanks to the type check.}


\subsection{Computability Properties}

We start with a few elementary properties stated without proofs:

\begin{lemma}
\label{l:eqcand}
Assume $\sigma\eqS\tau$. Then, $\cand{\sigma}=\cand{\tau}$.
\end{lemma}

\begin{lemma}
\label{l:arrowcomp}
$s=@(u,v)$ is computable if $u$ and $v$ are computable.
\end{lemma}

\begin{lemma}
\label{l:mts}
$s$ is computable if $s\in\sgs$ is strongly normalizable.
\end{lemma}

\begin{lemma}
\label{l:accessibility} 
Assume that $\vect{s}$ are computable and that $f(\vect{s})\gtacc v$
for some $f\in\F\cup\{@\}$. Then $v$ is computable.
\end{lemma}


We now give the fundamental properties of the interpretations. Note
that we use our term categorisation to define the computability
predicates, and that this is reflected in the computability properties
below.

(i) Every computable term is strongly normalizable for $\gtchorpoXtype{}$; 

(ii) If $s$ is computable and $s\gechorpoXtype{} t$, then
$t$ is computable;

(iii) A neutral term $s$ is computable iff $t$ is computable for
all terms $t$ such that $s\gtchorpoXtype{} t$;

(iv) An abstraction $\lambdax : \sigma.u$ is computable iff $u\{x\mapsto w\}$ is
computable for all computable terms $w:_C \sigma$;

(v) A prealgebraic term $s=f(\vect{s}):_C \sigma$ such that
$f:\vect{\sigma}\ra\tau\in\F$ is computable if
$\vect{s}:_C\vect{\sigma}$ is computable.

All proofs are adapted from~\cite{jouannaud06jacm}, with some
additional difficulties. The first four properties are proved
together.

\begin{proof} Properties (i), (ii), (iii), (iv).
Note first that the only if part of properties (iii) and (iv)
is property (ii). We are left with (i), (ii) and the if parts of (iii) and (iv)
which spell out as follows:

Given a type $\sigma$, we prove by induction on the definition
of $\cand{\sigma}$ that

(i) Given $s:_C\sigma\in\cand{\sigma}$, then $s$ is strongly
normalizable;

(ii) Given $s:_C\sigma\in\cand{\sigma}$ such that $s\gechorpoXtype{}
t$ for some $t:_C\tau$, then $t\in\cand{\tau}$;

(iii) A neutral candidate term $u:_C\sigma$ is computable if
$w:_C\theta\in\cand{\theta}$ for all $w$ such that $u\gtchorpoXtype{} w$;
in particular, variables are computable;

(iv) An abstraction $\lambda x:\alpha.u :_C \sigma$ is computable if
 $u\{x\mapsto w\}$ is computable for all $w\in\cand{\alpha}$.

We prove each property in turn, distinguishing in each case whether
$\sigma$ is a data or functional type.


\begin{itemize}
\item[(ii)]
\begin{enumerate}
\item
Assume that $\sigma$ is a data type.
The result holds by definition of the candidate interpretations.
\item Let $\sigma=\theta\ra\rho$.  By arrow
preservation and decreasingness properties, there are two cases:
\begin{enumerate}
\item $\rho \geS \tau$.  Let $y:_C\theta\in\X$. By induction
hypothesis (iii), $y\in\cand{\theta}$, hence $@(s,y)\in\cand{\rho}$ by
definition of $\cand{\sigma}$.  Since $@(s,y):_C\rho \gtchorpoXtype{}
t:_C\tau$ by case~\ref{appsubt} of the definition, $t$ is computable
by induction hypothesis (ii).
\item
$\tau=\theta'\ra\rho'$, with
$\theta\eqS\theta'$ and $\rho\geS\rho'$.  Since $s$ is computable,
given $u\in\cand{\theta}$, then $@(s,u)\in\cand{\rho}$.  By
monotonicity, $@(s,u)\gtchorpoXtype{X}@(t,u)$.  By
induction hypothesis (ii) $@(t,u)\in\cand{\rho'}$. Since
$\cand{\theta}=\cand{\theta'}$ by Lemma~\ref{l:eqcand}, $t$ is
computable by definition of $\cand{\tau}$.
\end{enumerate}
\end{enumerate}


\item[(i)]
\begin{enumerate}
\item
Assume first that $\sigma$ is a data type.  Let $s\gtchorpoXtype{}t$.
By definition of $\cand{\sigma}$, $t$ is computable, hence is strongly
normalizable by induction hypothesis. It follows $s$ is strongly
normalizable in this case.
\item
Assume now that $\sigma=\theta\ra\tau$, and let
$s_0=s:_C\sigma=\sigma_0 \gtchorpoXtype{}
s_1:_C\sigma_1\ldots\gtchorpoXtype{}
s_n:_C\sigma_n\gtchorpoXtype{}\ldots$ be a derivation issuing from
$s$. Therefore $s_i\in\cand{\sigma_i}$ by
induction on $i$, using the assumption that $s$ is computable for
$i=0$ and otherwise by the already proved property (ii).  Such
derivations are of the following two kinds:
  \begin{enumerate}
  \item
  $\sigma\gtS\sigma_i$ for some $i$, 
in which case $s_i$ is strongly
 normalizable by induction hypothesis (i).  The derivation
 issuing from $s$ is therefore finite.
  \item 
  $\sigma_i\eqS\sigma$ for all $i$, in which case 
$\sigma_i=\theta_i\ra\tau_i$ with $\theta_i\eqS\theta$. Then,
  $\{@(s_i, y:_C\theta):_C\tau_i\}_i$
  is a sequence of candidate terms  which is strictly
  decreasing with respect to $\gtchorpoXtype{}$ by monotonicity.
  Since $y:_C\theta$ is computable by induction hypothesis (iii),
  $@(s_i,y)$ is computable by definition of $\cand{\tau_i}$. By induction
  hypothesis, the above sequence is finite, implying that the
  starting sequence itself is finite. 
  \end{enumerate}
Therefore, $s$ is strongly normalizing as well in this case.
\end{enumerate}


\item[(iii)]
\begin{enumerate}
\item
Assume that $\sigma$ is a data type.
The result holds by definition of $\cand{\sigma}$.
\item
Assume now that $\sigma=\sigma_1\ra\sigma_2$. By definition of
$\cand{\sigma}$, $u$ is computable if the neutral term $@(u,u_1)$ is
computable for all $u_1\in\cand{\sigma_1}$. By induction hypothesis,
$@(u,u_1)$ is computable iff all its reducts $w$ are computable.

Since $u_1$ is strongly normalizable by induction hypothesis (i), we
show by induction on the pair $(u_1,|w|)$ ordered by
$(\gtchorpoXtype{},>_{\Nat})$ that all reducts $w$ of $@(u,u_1)$ are
computable.  Since $u$ is neutral, hence is not an abstraction, there
are three possible cases:
\begin{enumerate}
\item
$@(u,u_1)\gtchorpoXtype{} w$ by Case~\ref{appsubt}, therefore
$u\geacc v\gechorpoXtype{} w$ or $u_1\geacc v\gechorpoXtype{} w$ for some
$v$. Since the type of $w$ is smaller or equal to the type of
$@(u,u_1)$, it is strictly smaller than the type of $u$, hence $w\neq
u$. Therefore, in case $v=u$, $w$ is a reduct of $u$, hence is
computable by assumption.  Otherwise, $v$ is $u_1$ or a minimal-type
subterm of $u_1$, in which case it is computable by assumption on
$u_1$ and Lemma~\ref{l:mts}, or a minimal-type subterm of $u$ in which
case $u\gtchorpoXtype{} v$ by Case~\ref{subt} or~\ref{appsubt} since
the neutral term $u$ is not an abstraction, and therefore $v$ is
computable by assumption. It follows that $w$ is computable by
induction hypothesis (ii).
\item
$@(u,u_1)\gtchorpoXtype{} w$ by Case~\ref{appstat}, therefore
$w=@(v,v_1)$ and also $\{u,u_1\}
(\gechorpoXtype{})_{mul}\{w_1,w_2\}$.  For type reasons, there are
again two cases:
\begin{itemize}
\item
$w_1$ and $w_2$ are strictly smaller than $u,u_1$, in which case
 $w_1$ and $w_2$ are computable by assumption or induction
 hypothesis (ii), hence $w$ is computable by Lemma~\ref{l:arrowcomp}.
\item
 $u=w_1$ and $u_1 \gtchorpoXtype{} w_2$, implying that $w_2$ is
 computable by assumption and induction hypothesis (ii).  Then, since 
$(u_1,\_)
 (\gtchorpoXtype{},>_{\Nat})_{lex} (w_2,\_)$, we 
 conclude by induction hypothesis.
\end{itemize}

\item
$@(u,u_1)\gtchorpoXtype{} w$ by Case~\ref{appabs}, then
$w=\lambdax:\beta.w'$, $x\not\in\Var{w'}$ and $@(u,u_1)\gtchorpoX{}
w'$. By induction hypothesis (iv) and the fact that
$x\not\in\Var{w'}$, $w$ is computable if $w'$ is computable.  Since
the type of $\lambdax:\beta.w'$ is strictly bigger than the type of
$w'$, we get $@(u,u_1)\gtchorpoXtype{} w'$. We conclude by induction
hypothesis, since $(u_1,\lambdax.w') (\gtchorpoXtype{},>_{\Nat})_{lex}
(u_1,w')$.

\end{enumerate}
\end{enumerate}


\item[(iv)] By definition of $\cand{\sigma}$, the abstraction
$\lambdax:\alpha.u :_C\sigma$ is computable if the term
$@(\lambdax.u,w)$ is computable for an arbitrary $w\in\cand{\alpha}$.

Since variables are computable by induction hypothesis (iii),
$u=u\{x\mapsto x\}$ is computable by assumption. By induction
hypothesis (i), $u$ and $w$ are strongly normalizable.
We therefore
prove that $@(\lambdax.u,w)$ is computable by induction on the pair
$(u,w)$ compared in the ordering
$(\gtchorpoXtype{},\gtchorpoXtype{})_{lex}$. 

Since $@(\lambdax.u,w)$ is neutral, we need to show that all reducts
 $v$ of $@(\lambdax.u,w)$ are computable. We consider the four possible cases
 in turn:
\begin{enumerate}

\item
If $@(\lambdax.u,w)\gtchorpoXtype{}v$ by Case~\ref{appsubt}, there are two cases: 

- if $w\gechorpoXtype{} v$, we conclude by induction hypothesis (ii) that
$v$ is computable.

- if $\lambdax.u\gechorpoXtype{} v$, then $\lambdax.u\gtchorpoXtype{}
v$ since the type of $\lambdax.u$ must be strictly bigger than the
type of $v$.  There are two cases depending on the latter
comparison.

If the comparison is by Case~\ref{abssubt}, then
$u\gechorpoXtype{}v$, and we conclude by induction hypothesis (ii)
that $v$ is computable.

If the comparison is by Case~\ref{absstat}, then
$v=\lambdax:\alpha'.u'$ with $\alpha\eqS\alpha'$.  By stability,
$u\{x\mapsto w\}\gtchorpoXtype{} u'\{x\mapsto w\}$, hence
$u'\{x\mapsto w\}$ is computable by property (ii) for an arbitrary
$w\in\cand{\alpha}=\cand{\alpha'}$ by lemma~\ref{l:eqcand}. It follows
that $v$ is computable by induction hypothesis, since
$(u,\_) (\gtchorpoXtype{},\gtchorpoXtype{})_{lex} (u',\_)$.

\item
If $@(\lambdax.u,w)\gtchorpoXtype{}v$ by case~\ref{appstat}, then
$v=@(v_1,v_2)$, and by definition of $\gtchorpoX{}$,
$\{\lambdax.u,w\} (\gtchorpoXtype{})_{mul}\{v_1,v_2\}$. There are three cases:

- $v_1=\lambdax.u$ and $w\gtchorpoXtype{} v_2$. Then $v_2$ is
computable by induction hypothesis (ii) and, since $u\{x\mapsto v_2\}$
is computable by the main assumption, $@(v_1,v_2)$ is computable by
induction hypothesis, since $(\lambdax.u,w)
(\gtchorpoXtype{},\gtchorpoXtype{})_{lex}(\lambdax.u,v_2)$.

- Terms in $\{v_1,v_2\}$ are reducts of $u$ and $w$. Therefore,
  $v_1$ and $v_2$ are computable by induction hypothesis (ii) and $v$
is computable by Lemma~\ref{l:arrowcomp}.

- Otherwise, for typing reason, $v_1$ is a reduct of $\lambdax.u$ of
the form $\lambdax.u'$ with $u\gtchorpoXtype{} u'$, and $v_2$ is a
reduct of the previous kind.  By the main assumption, $u\{x\mapsto
v''\}$ is computable for an arbitrary computable $v''$.  Besides,
$u\{x\mapsto v''\}\gtchorpoXtype{} u'\{x\mapsto v''\}$ by stability.
Therefore $u'\{x\mapsto v''\}$ is computable for an arbitrary
computable $v''$ by induction hypothesis (ii).  Then $@(v_1,v_2)$ is
computable by induction hypothesis, since $(u,\_)$
$(\gtchorpoXtype{},\gtchorpoXtype{})_{lex}$ $(u',\_)$.

\item
If $@(\lambdax.u,w)\gtchorpoXtype{}v$ by Case~\ref{appabs}, then
$v=\lambda x.v'$, $x\not\in\Var{v'}$ and
$@(\lambdax.u,w)\gtchorpoXtype{}v'$. Since
$\lambdax.v'\gtchorpoXtype{}v'$ by Case~\ref{abssubt}, $v'$ is
computable by induction hypothesis.  Since $x\not\in\Var{v'}$, it
follows that $\lambdax.v'$ is computable.

\item
If $@(\lambdax.u,w)\gtchorpoXtype{} v$ by case~\ref{beta},
then $u\{x\mapsto w\}\gehorpo v$. By assumption, $u\{x\mapsto w\}$ is
computable, and hence $v$ is computable by property (ii).
\cqfd
\end{enumerate}
\end{itemize}
\end{proof}


We are left with property (v) whose proof differs from~\cite{jouannaud06jacm}.

\begin{proof}
Property (v). As we have seen, each data type interpretation
$\cand{\sigma}$ is the least fixpoint of a monotone function $G$ on
the powerset of the set of terms. Hence, for every computable term
$t\in\cand{\sigma}$, there exists a smallest ordinal $o(t)$ such that
$t\in G^{o(t)}(\emptyset)$, where $G^a$ is the $a$ transfinite
iteration of $G$. The relation $\qgt$, defined by $t\qgt u$ iff
$o(t)>o(u)$, is a well-founded ordering which is compatible with
$\gtchorpoXtype{}$: if $t\gtchorpoXtype{}u$ then $t\qge u$. The proof
is by induction on the type ordering. Therefore,
${\gtchorpoXtype{}}\cup{\qgt}$ is well-founded on computable
terms. Note that the result would again hold for terms headed by a
function symbol with a functional output.

We use this remark to build our outer induction argument: we prove
that $f(\vect{s})$ is computable by induction on the pair
$(f,\vect{s})$ ordered lexicographically by
$(\gtF,(\gtchorpoXtype{}\cup\qgt)_{stat_f})_{lex}$. This is our outer
statement (OH).
  
  Since $f(\vect{s})$ is prealgebraic, it is computable
  if every subterm at an accessible position is computable (which
  follows by assumption) and reducts $t$ of $s$ are computable.

Since $\gtchorpoXtype{}$ is defined in terms of $\gtchorpoX{X}$, we
  actually prove by an inner induction on the recursive definition of
  $\gtchorpoX{X}$ the more general inner statement (IH) that $t\gamma$
  is computable for an arbitrary term $t$ such that
  $f(\vect{s})\gtchorpoX{X} t$ and computable substitution $\gamma$ of
  domain $X$ such that $X\cap\Var{s}=\emptyset$.  Since the identity
  substitution is computable by property (iii), our inner induction
  hypothesis implies our outer induction hypothesis.

\begin{enumerate}
\item
If $f(\vect{s})\gtchorpoX{X} u$ by Case~\ref{var}, Then $u\in X$ and
we conclude by assumption on $\gamma$ that $u\gamma$ is computable.

\item 
If $f(\vect{s})\gtchorpoX{X} u$ by Case~\ref{subt}, then $s_i\geacc t$
for some $i$ and $@(t,\vect{x})\gechorpoXtype{} u$ for some
$\vect{x}\in X$. By assumption on $\vect{s}$ and
Lemma~\ref{l:accessibility}, $t$ is computable. Since $t$ is a subterm
of $s$ and $X\cap\Var{s}=\emptyset$, then $t\gamma=t$ is
computable. It follows that $@(t,\vect{x}\gamma)$ is computable. Thus,
by stability, $u\gamma$ is computable.

\item If $f(\vect{s})\gtchorpoX{X} u$ by case~\ref{stat}, then
$u=g(\vect{u})$, $f=_\F g$, $s\gtchorpoX{X} \vect{u}$ and finally
$\vect{s}~({\gtchorpoXtype{}}\cup{\gtchorpoXtypeacc{X}})_{stat}~\vect{u}$.
By the inner induction hypothesis, $\vect{u}\gamma$ is computable.
Assume now that $s_i:\sigma_i\gtacc v$ and
$@(v,\vect{x}):\sigma_i'\eqS\sigma_i\gechorpoXtype{}u_j$. Using the
fact that $X\cap\Var{s}=\emptyset$, by stability we get
$s_i\gamma=s_i{\gtacc} v\gamma=v$ and
$@(v,\vect{x})\gamma=@(v,\vect{x}\gamma):\sigma_i'\eqS\sigma_i\gechorpoXtype{}u_j\gamma$. Moreover,
by definition of computability, $s_i\qgt@(v,\vect{x}\gamma)$.
Therefore, $u\gamma=f(\vect{u}\gamma)$ is computable by the outer
induction hypothesis.

\item
If $f(\vect{s})\gtchorpoXtype{X} u$ by case~\ref{precabs}, then
$u=\lambdax.v$ with $x\not\in\Var{s}$ and
$f(\vect{s})\gtchorpoX{X\cup\{x\}} v$. By the inner induction
hypothesis, $v(\gamma\cup\{x\mapsto w\})$ is computable for an
arbitrary computable $w$. Assuming without loss of generality that
$x\not\in\Ran{\gamma}$, then $v(\gamma\cup\{x\mapsto
w\})=(v\gamma)\{x\mapsto w\}$.  Therefore, $u=\lambdax.v\gamma$ is
computable by computability property (iv).

\item
If $f(\vect{s})\gtchorpoX{X} u$ by Case~\ref{prec}, then
$u=g(\vect{u})$ with $g\in\F\cup\{@\}$ and
$s\gtchorpoX{X}\vect{u}$. By the inner induction hypothesis,
$\vect{u}\gamma$ is computable. We conclude by Lemma~\ref{l:arrowcomp}
in case $g=@$ and by the outer induction hypothesis if $g\in\F$.
\cqfd
\end{enumerate}
\end{proof}


\subsection{Strong Normalization}

We are now ready for the strong normalization proof. From the previous
properties, one can easily prove the following lemma by induction on
the term structure:

\begin{lemma}
\label{main}
Let $\gamma$ be a type-preserving computable substitution and $t$ be an algebraic
$\lambda$-term. Then $t\gamma$ is computable.
\end{lemma}

\comment{\begin{proof}
  The proof proceeds by induction on the size of $t$.
\begin{enumerate}
\item
  $t$ is a variable $x$.  Then $x\gamma$ is computable by assumption.
\item $t$ is an abstraction $\lambdax.u$. By computability property
  (v), $t\gamma$ is computable if $u\gamma\{x\mapsto w\}$ is
  computable for every well-typed computable candidate term
  $w$. Taking $\delta= \gamma\cup\{x\mapsto w\}$, we have
  $u\gamma\{x\mapsto w\}=u(\gamma\cup\{x\mapsto w\})$ since $x$ may
  not occur in $\gamma$.  Since $\delta$ is computable and $|t| >
  |u|$, by induction hypothesis, $u\delta$ is computable.
\item $t=@(t_1,t_2)$. Then $t_1\gamma$ and $t_2\gamma$ are computable
by induction hypothesis, hence $t$ is computable by
Lemma~\ref{l:arrowcomp}.
\item $t=f(t_1,\ldots,t_n)$.
Then $t_i\gamma$ is computable
by induction hypothesis, hence
$t\gamma$ is computable by computability property (v).
 \qed
\end{enumerate}
\end{proof}}

The proof of our main theorem follows from Lemma~\ref{main} when using
the identity substitution, and of computability property (i).

%% file: conclusion.tex
\section{Conclusion}
\label{s:conclusion}

An implementation of the new definition with examples is available
from the web page of the third author.

There are still a few possible improvements that we have not yet
explored, such as ordering the abstractions according to their type,
increasing the set of accessible terms for applications that satisfy
the strict positivity restriction, and showing that the new definition
is strictly more general that the general schema when adopting the
same type discipline. A more difficult problem to be investigated then
is the generalization of this new definition to the calculus of
constructions along the lines of~\cite{walukiewicz00lfm}.

%% file: main.bbl
\begin{thebibliography}{1}

\bibitem{blanqui06rr}
F.~Blanqui.
\newblock {(HO)RPO} revisited.
\newblock Research Report 5972, INRIA, 2006.

\bibitem{jouannaud06lpar}
F.~Blanqui, J.-P. Jouannaud, and A.~Rubio.
\newblock Higher order termination: from {K}ruskal to computability.
\newblock In {\em Proc. LPAR, Phnom Penh, Cambodgia, LNCS 4246}, 2006.

\bibitem{jouannaud99lics}
Jean-Pierre Jouannaud and Albert Rubio.
\newblock The higher-order recursive path ordering.
\newblock In {\em 14th {IEEE} Symposium on Logic in Computer Science}, 1999.

\bibitem{jouannaud06jacm}
Jean-Pierre Jouannaud and Albert Rubio.
\newblock Polymorphic higher-order recursive path orderings.
\newblock {\em Journal of the ACM}, 54(1):1--48, 2007.

\bibitem{walukiewicz00lfm}
Daria Walukiewicz-Chrzaszcz.
\newblock Termination of rewriting in the {C}alculus of {C}onstructions.
\newblock In {\em Proceedings of the Workshop on Logical Frameworks and
  Meta-languages\comment{, {S}anta {B}arbara, {C}alifornia}}, 2000.

\end{thebibliography}
